\documentclass[prl,preprint,showpacs,showkeys]{revtex4}
\usepackage{graphicx}
 \RequirePackage{times}
 \RequirePackage{mathptm}
\begin{document}


\title{Finding a state in a haystack}
\author{Niko Donath and Karl Svozil}
 \email{svozil@tuwien.ac.at}
\homepage{http://tph.tuwien.ac.at/~svozil}
\affiliation{Institut f\"ur Theoretische Physik, University of Technology Vienna,
Wiedner Hauptstra\ss e 8-10/136, A-1040 Vienna, Austria}

\begin{abstract}
We consider the problem to single out a
particular state among $2^n$ orthogonal pure states.
As it turns out, in general the optimal strategy is not to measure the particles separately,
but to consider joint properties of the $n$-particle system.
The required number of propositions is $n$.
There exist $2^n!$ equivalent operational procedures to do so.
We enumerate some configurations for three particles,
in particular the Greenberger-Horne-Zeilinger (GHZ)- and W-states,
which are specific cases of a unitary transformation
For the GHZ-case,
an explicit physical meaning of the projection operators is discussed.
\end{abstract}

\pacs{03.65.Ta,03.65.Fd03.67.-a}
\keywords{GHZ-states,W-states,quantum measurement theory}

\maketitle


Suppose ``Bob'' is told that ``Alice'' has prepared  $n$ two-state systems
in a particular pure state  among $N=2^n$ pure states.
Assume further that these pure states  correspond to a complete
orthonormal basis of some $N$-dimensional Hilbert space.
Bob's task is to find out which particular single one of the ``haystack'' of $2^n$
states Alice has chosen to communicate to him
\footnote{
A related classical question with different solutions is the problem of ascerting
the minimum number of weighings which suffice to determine $k$ defective
coins in a set of $N$ coins of the same appearence, given an equal arm balance
\cite{smith47,cairns63}.}.

As can be expected, there exist efficient and expensive
search strategies for such a task.
The ``worst'' strategy (besides mere repetition)
would be to check the proposition corresponding to each individual pure state,
asking, ``is the system in state $i$?'' for $i=1,\ldots ,2^n$.
This strategy would require $2^{n/2}$ questions on the average and could take
$2^n$ questions at worst until one reaches a positive answer.
Yet, by exploiting the {\em joint properties} of the $n$ systems,
we may expect to reduce the complexity of Bob's task.

In what follows we shall thus deal with the following questions
aimed at a systematic understanding of the ``haystack'' problem.
(i) What is the minimal set of propositions (i.e., operationalizable yes-no statements)
which singles out a  particular
pure state of $n$ entangled two-state systems from other orthogonal pure states?
(ii) How many different but equivalent sets of propositions can be defined?
(iii) What is the explicit form and physical interpretation of the propositions associated with an arbitrary basis?

As it turns out, the number of propositions required for solving the  ``haystack problem''
can be reduced to $n$, which is an exponential gain with respect to
the worst strategy just mentioned.
This result is in agreement with
Zeilinger's foundational principle stating that
$n$ elementary two-state systems carry $n$ classical bits; i.e., the answer to
at most $n$ questions concerning their physical properties \cite{zeil-99}.
In classical information theory a proposition is the yes-no
statement settling a question with two possible answers.
In standard quantum logic  \cite{birkhoff-36,ma-57,kochen1,svozil-ql},
quantum propositions are identified with projection operators.
The eigenvalues 0 and 1 of these projection operators
are identified with the two possible yes-no answers, respectively.

Conversely, by assuming Zeilinger's principle,
it should be possible to define $n$-particle quantum states by
the set of eigenvalues associated with quantum mechanical propositions.
When choosing the ``optimal'' strategy, $n$ propositions should suffice.
However, one could also take an arbitrary number of consistent propositions.
If these are not ``optimal'' in a well-defined sense,
then this results in nonpure quantum states.



Consider a $2^n=N$--dimensional
Hilbert space of $n$ particles in two states (labeled by ``$1$,'' ``$2$,''
or ``up,''``down,'' or ``$+$,'' ``$-$,'' or whatever).
The standard orthonormal (``Cartesian'') basis is given by
(superscript ``T'' indicates transposition)
\begin{eqnarray}
\{|e_1\rangle = |+++\cdots +\rangle &\equiv& (1,0,0,\ldots ,0)^T,
\nonumber \\
|e_2\rangle = |+++\cdots -\rangle &\equiv& (0,1,0,\ldots ,0)^T,
\nonumber \\
&\vdots&
\nonumber \\
|e_N\rangle = |---\cdots -\rangle &\equiv& (0,0,0,\ldots ,1)^T\}.
\label{2001-principle-scb}
\end{eqnarray}

Let us first concentrate on an enumeration of all propositions which uniquely distinguish
the $N$ vectors that form an orthogonal basis of an $N$-dimensional vector space.
This task corresponds
to the construction of projection operators---which have some operational interpretation(s)
in terms of (quantum mechanical) measurements---whose combined effect is
the separation of each individual base state from all the other ones.
In that respect, the measurement apparatus and the associated propositions
act as filters which effectively generate
a {\em partitioning} of some orthonormal basis into
partitions which contain only single elements of that basis
(i.e., one basis element per partition element).
Formally, the projections induce an equivalence relation on the set of base states.

We shall impose the following requirements upon the propositions.
(i) All propositions are co-measurable (i.e., the associated projections commute).
(ii) Any single proposition separates half of the elements of the orthogonal base vectors
from the other half;
i.e., any proposition $F_i$ generates a $50:50$ partition $f_i$ with $\vert f_i\vert =2$
and $\vert f_{i,k}\vert = N/2$
(``$\vert x \vert $'' stands for the number of elements of a finite set $x$),
$f_{i,k}\in f_i$, $k=1,2$.
(iii) For any two propositions $F_i,F_j$, $i\neq j$, the intersection
of elements of the associated partitions $f_i,f_j$ of some
orthogonal basis reduce the
size of the elements of the partitions by a factor of two.

We shall introduce an optimal algorithm implementing these requirements
which uses exactly $n$ propositions to decompose every orthonormal basis of
the $N$-dimensional vector space.
It implements a binary search strategy which can be enumerated as follows:
separate the first ${N / 2}$ vectors from the second ${N / 2}$.
Then, within every such block, separate the first ${N / 4}$ vectors from the second ${N / 4}$.
Iterate these procedure by reducing the block size by a factor of two in each step until
blocks of size one are reached.
This ``state sieve'' is an optimal search strategy in the sense that
in general no shorter proposition system
exists which  separates each individual state of the
standard orthonormal basis.

The explicit form of the operators are
(``$\textrm{diag}$'' stands for a diagonal matrix)
\begin{eqnarray}
O_1&=&\textrm{diag}\left(\underbrace{1,\ldots ,1}_{N/2},\underbrace{0,\ldots,0}_{N/2}\right),\\
O_2&=&\textrm{diag}\left(\underbrace{1,\ldots ,1}_{N/4},\underbrace{0,\ldots,0}_{N/4},\underbrace{1,\ldots ,1}_{N/4}),\underbrace{0,\ldots,0}_{N/4}\right),\\
&\vdots &\\
O_n&=&\textrm{diag}\left(\underbrace{1,0,1,0,\ldots ,1,0}_{N}\right),
\end{eqnarray}
and the orthogonal operators $O_i' =1-O_i$. All these projections commute with one another.
Their associated propositions can be stated as follows.\\
$O_1\equiv $ ``The first particle is in state $+$.''     \\
$O_2\equiv $ ``The second particle is in state $+$.''      \\
$\vdots$                                                     \\
$O_n\equiv $ ``The $n$th particle is in state $+$.''

It is easy to verify that these operators are projection operators and that they mutually commute; i.e.,
$O_iO_i=O_i,\;  \big[O_i,O_j\big]=0,$ for all $i,j\in \{1,\ldots ,N\}, i\neq j$.

Let us now turn to the question of how many equivalent systems
of operators and propositions exist.
Notice that only diagonal matrices containing $1$s and $0$s
in the principal diagonal can be eigenmatrices of the standard
orthonormal basis (\ref{2001-principle-scb}).
For this particular basis, the only possible variations are obtained as follows.
First, the diagonals of $O_1,O_2,\ldots ,O_n$ are written below each other.
If one considers the columns of this listing,
each one of the $N$ columns of length $n$ represents a unique number
$N= a_1> a_2> \cdots > a_{N-1}> a_N =0$
in binary notation and in strictly decreasing order.
Other valid state sieves are obtained by exchanging two arbitrary columns.
This amounts to the permutation of $N$ different $n$-ary columns.
The total number of such entities and thus of all equivalent systems of $n$ propositions
is $N!=2^n!\;$.

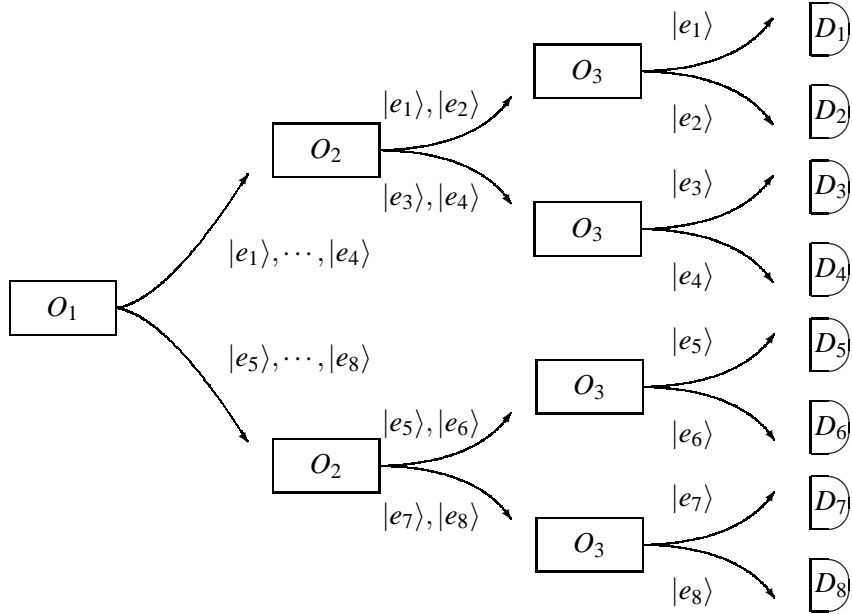
\begin{figure}
\begin{center}
\unitlength 0.70mm
\linethickness{0.4pt}
\begin{picture}(159.34,115.67)
\put(0.00,52.67){\framebox(20.00,10.00)[cc]{$O_1$}}
\put(50.00,82.67){\framebox(20.00,10.00)[cc]{$O_2$}}
\put(95.00,97.67){\vector(2,3){0.2}}
\bezier{120}(70.00,87.67)(88.33,88.00)(95.00,97.67)
\put(95.00,77.67){\vector(2,-3){0.2}}
\bezier{120}(70.00,87.67)(88.33,87.34)(95.00,77.67)
\put(50.00,22.67){\framebox(20.00,10.00)[cc]{$O_2$}}
\put(95.00,37.67){\vector(2,3){0.2}}
\bezier{120}(70.00,27.67)(88.33,28.00)(95.00,37.67)
\put(95.00,17.67){\vector(2,-3){0.2}}
\bezier{120}(70.00,27.67)(88.33,27.34)(95.00,17.67)
\put(100.00,37.67){\framebox(20.00,10.00)[cc]{$O_3$}}
\put(145.00,52.67){\vector(2,3){0.2}}
\bezier{120}(120.00,42.67)(138.33,43.00)(145.00,52.67)
\put(145.00,32.67){\vector(2,-3){0.2}}
\bezier{120}(120.00,42.67)(138.33,42.34)(145.00,32.67)
\put(100.00,7.67){\framebox(20.00,10.00)[cc]{$O_3$}}
\put(145.00,22.67){\vector(2,3){0.2}}
\bezier{120}(120.00,12.67)(138.33,13.00)(145.00,22.67)
\put(145.00,2.67){\vector(2,-3){0.2}}
\bezier{120}(120.00,12.67)(138.33,12.34)(145.00,2.67)
\put(99.67,97.67){\framebox(20.00,10.00)[cc]{$O_3$}}
\put(144.67,112.67){\vector(2,3){0.2}}
\bezier{120}(119.67,102.67)(138.00,103.00)(144.67,112.67)
\put(144.67,92.67){\vector(2,-3){0.2}}
\bezier{120}(119.67,102.67)(138.00,102.34)(144.67,92.67)
\put(99.67,67.67){\framebox(20.00,10.00)[cc]{$O_3$}}
\put(144.67,82.67){\vector(2,3){0.2}}
\bezier{120}(119.67,72.67)(138.00,73.00)(144.67,82.67)
\put(144.67,62.67){\vector(2,-3){0.2}}
\bezier{120}(119.67,72.67)(138.00,72.34)(144.67,62.67)
\put(45.00,83.00){\vector(2,3){0.2}}
\bezier{152}(20.00,57.67)(29.67,59.00)(45.00,83.00)
\put(45.00,32.34){\vector(2,-3){0.2}}
\bezier{152}(20.00,57.67)(29.67,56.34)(45.00,32.34)
\put(152.17,5.00){\oval(14.33,10.00)[r]}
\put(152.00,-0.00){\line(0,1){10.00}}
\put(156.00,5.00){\makebox(0,0)[cc]{$D_8$}}
\put(152.17,20.67){\oval(14.33,10.00)[r]}
\put(152.00,15.67){\line(0,1){10.00}}
\put(156.00,20.67){\makebox(0,0)[cc]{$D_7$}}
\put(152.17,35.00){\oval(14.33,10.00)[r]}
\put(152.16,65.00){\oval(14.33,10.00)[r]}
\put(152.17,95.00){\oval(14.33,10.00)[r]}
\put(152.00,30.00){\line(0,1){10.00}}
\put(152.00,60.00){\line(0,1){10.00}}
\put(152.00,90.00){\line(0,1){10.00}}
\put(156.00,35.00){\makebox(0,0)[cc]{$D_6$}}
\put(156.00,65.00){\makebox(0,0)[cc]{$D_4$}}
\put(156.00,95.00){\makebox(0,0)[cc]{$D_2$}}
\put(152.17,50.67){\oval(14.33,10.00)[r]}
\put(152.16,80.67){\oval(14.33,10.00)[r]}
\put(152.17,110.67){\oval(14.33,10.00)[r]}
\put(152.00,45.67){\line(0,1){10.00}}
\put(152.00,75.67){\line(0,1){10.00}}
\put(152.00,105.67){\line(0,1){10.00}}
\put(156.00,50.67){\makebox(0,0)[cc]{$D_5$}}
\put(156.00,80.67){\makebox(0,0)[cc]{$D_3$}}
\put(156.00,110.67){\makebox(0,0)[cc]{$D_1$}}
\put(55.00,68.00){\makebox(0,0)[cc]{$\vert e_1\rangle ,\cdots ,\vert e_4\rangle$}}
\put(55.00,48.00){\makebox(0,0)[cc]{$\vert e_5\rangle ,\cdots ,\vert e_8\rangle$}}
\put(80.00,96.33){\makebox(0,0)[cc]{$\vert e_1 \rangle ,\vert e_2 \rangle $}}
\put(80.00,78.67){\makebox(0,0)[cc]{$\vert e_3 \rangle ,\vert e_4 \rangle $}}
\put(80.00,35.67){\makebox(0,0)[cc]{$\vert e_5 \rangle ,\vert e_6 \rangle $}}
\put(80.00,18.00){\makebox(0,0)[cc]{$\vert e_7 \rangle ,\vert e_8 \rangle $}}
\put(129.67,111.33){\makebox(0,0)[cc]{$\vert e_1 \rangle $}}
\put(129.67,93.67){\makebox(0,0)[cc]{$\vert e_2 \rangle $}}
\put(129.67,81.33){\makebox(0,0)[cc]{$\vert e_3 \rangle $}}
\put(129.67,51.33){\makebox(0,0)[cc]{$\vert e_5 \rangle $}}
\put(129.67,21.33){\makebox(0,0)[cc]{$\vert e_7 \rangle $}}
\put(129.67,63.67){\makebox(0,0)[cc]{$\vert e_4 \rangle $}}
\put(129.67,33.67){\makebox(0,0)[cc]{$\vert e_6 \rangle $}}
\put(129.67,3.67){\makebox(0,0)[cc]{$\vert e_8 \rangle $}}
\end{picture}
\end{center}
\caption{State sieve resulting from binary search.
Successive measurements of propositions $O_1,O_2,O_3$ serve as filters to
single out the input state. $D_1,\ldots ,D_8$ indicate the final detectors;
In the lossless case, exactly one of them clicks.}
\label{2000-principle-f1}
\end{figure}

Let us demonstrate the construction by considering the case
$n=3, N=8$ (e.g., three spin $1/2$ particles).
The operators can be written in a diagonal form
\begin{eqnarray}
O_1&=&\textrm{diag}\left(1,1,1,1,0,0,0,0\right), \label{2001-principle-e1}\\
O_2&=&\textrm{diag}\left(1,1,0,0,1,1,0,0\right),  \\
O_3&=&\textrm{diag}\left(1,0,1,0,1,0,1,0\right). \label{2001-principle-e3}
\end{eqnarray}
Some of the permutations yielding $8!=40320$ equivalent systems of three propositions
are enumerated in Table \ref{2001-principle-tc}.
\begin{table}
\begin{center}
\begin{tabular}{rrrrrrrrrrrr}

\begin{tabular}{rrrrrrrrrrrr}
\hline
11110000\\
11001100\\
10101010\\
\hline
\end{tabular}
&
$\longleftrightarrow$
\begin{tabular}{rrrrrrrrrrrr}
\hline
11110000\\
11001100\\
01101010\\
\hline
\end{tabular}
&
$\longleftrightarrow$
&
$\cdots$
&
$\longleftrightarrow$
\begin{tabular}{rrrrrrrrrrrr}
\hline
00001111\\
00110011\\
01010101\\
\hline
\end{tabular}
\end{tabular}
\end{center}
\caption{Enumeration of the $8!$ equivalent variations of propositions for $n=3,N=2^3=8$.
\label{2001-principle-tc}}
\end{table}
The cascade of filters
representable by projection operators
(\ref{2001-principle-e1})---(\ref{2001-principle-e3})
and interpretable as elementary yes-no propositions are depicted in
Figure \ref{2000-principle-f1}.
Any permutation of these measurements yields the same partitioning of states.

An arbitrary orthonormal basis of an $N$-dimensional vector space can be defined
as the isometric transforms  of the standard orthonormal basis (\ref{2001-principle-scb}).
If the vector space is complex (i.e., $C^N$),
these isometries are the unitary transforms.
Furthermore, any basis change in $C^N$ from one orthonormal basis into another one
can be represented by
some unitary matrix $U$; i.e., $|b_i\rangle=U |e_i\rangle $.
The group of unitary transformations $U(N)$ in $N$-dimensional
Hilbert space has $N^2$ parameters.
In the following we shall study this entire group rather than the transformations resulting
from the combined effect of $[U(2)]^n$, which are again unitary transformations.

Thus the problem of finding the $N$ propositions for the basis $|b_i\rangle$
can simply be solved by transforming the propositions for the vector space; i.e.,
$
O^b_i=U O^e_i U^{-1}
$.
Here, $O =O^e$ for some $O^e\in \{O_1,\ldots ,O_N\}$.
These propositions have the same eigenvalues as the  propositions, since if we identify
$O^e |e_i\rangle = \lambda |e_i \rangle $, then
$
O^b |b_i \rangle = U O^e U^{-1} U |e_i \rangle = \lambda U |e_i \rangle = \lambda |b_i \rangle
$.

In ref. \cite{zeil-99} Zeilinger poses the
following question, ``what are  the three propositions
which can be used to uniquely define the eight states of the three-particle case?''
Consider the eight orthonormal GHZ-basis states
\begin{eqnarray}
&&|\psi_1\rangle=\frac{1}{\sqrt{2}} \left( |+++\rangle+|---\rangle \right) \equiv (1,0,0,0,0,0,0,1)^T \equiv |111\rangle \nonumber \\
&&|\psi_2\rangle=\frac{1}{\sqrt{2}} \left( |+++\rangle-|---\rangle \right) \equiv (1,0,0,0,0,0,0,-1)^T\equiv |110\rangle \nonumber \\
&&|\psi_3\rangle=\frac{1}{\sqrt{2}} \left( |++-\rangle+|--+\rangle \right) \equiv (0,1,0,0,0,0,1,0)^T\equiv |101\rangle \nonumber \\
&&|\psi_4\rangle=\frac{1}{\sqrt{2}} \left( |++-\rangle-|--+\rangle \right) \equiv (0,1,0,0,0,0,-1,0)^T\equiv |011\rangle \nonumber \\
&&|\psi_5\rangle=\frac{1}{\sqrt{2}} \left( |+-+\rangle+|-+-\rangle \right) \equiv (0,0,1,0,0,1,0,0)^T\equiv |100\rangle \nonumber \\
&&|\psi_6\rangle=\frac{1}{\sqrt{2}} \left( |+-+\rangle-|-+-\rangle \right) \equiv (0,0,1,0,0,-1,0,0)^T\equiv |010\rangle \nonumber \\
&&|\psi_7\rangle=\frac{1}{\sqrt{2}} \left( |-++\rangle+|+--\rangle \right) \equiv (0,0,0,1,1,0,0,0)^T\equiv |001\rangle \nonumber \\
&&|\psi_8\rangle=\frac{1}{\sqrt{2}} \left( |-++\rangle-|+--\rangle \right) \equiv (0,0,0,1,-1,0,0,0)^T\equiv |000\rangle .\nonumber \\
\end{eqnarray}
They can be  interpreted as follows. The relative directions of the three spins are fixed
but their respective values undetermined. Thus one measurement on each one of the
three particles will suffice to know the exact
values of all spins.
It is possible to characterize the states according to the truth value of the propositions below
by $|111\rangle $ to $|000\rangle$.
Just as before, let us define the vector components of the standard orthonormal basis states
as
$|+++\rangle \equiv
(1,0,0,0,0,0,0,0)^T, \ldots ,
|---\rangle
\equiv
 (0,0,0,0,0,0,0,1)^T $.
The unitary matrix which transforms this standard basis
into the GHZ-basis is given by
\begin{equation}
U^{\textrm{GHZ}}={1 \over \sqrt{2}}\left(\begin{array}{rrrrrrrr}
1 & 1 & 0 & 0 & 0 & 0 & 0 & 0 \\
0 & 0 & 1 & 1 & 0 & 0 & 0 & 0 \\
0 & 0 & 0 & 0 & 1 & 1 & 0 & 0 \\
0 & 0 & 0 & 0 & 0 & 0 & 1 & 1 \\
0 & 0 & 0 & 0 & 0 & 0 & 1 & -1 \\
0 & 0 & 0 & 0 & 1 & -1 & 0 & 0 \\
0 & 0 & 1 & -1 & 0 & 0 & 0 & 0 \\
1 & -1 & 0 & 0 & 0 & 0 & 0 & 0
\end{array}\right)
.
\end{equation}
An explicit calculation shows that the matrices $O_i=O^{\textrm{GHZ}}_i=U^{\textrm{GHZ}} O^e_i {U^{\textrm{GHZ}}}^{-1}$, $i=1,2,3$,
can be written as follows.
\begin{eqnarray}
&&O_1=\textrm{diag}\left(1,1,0,0,0,0,1,1\right)\label{2001-principle-e11},\\
&&O_2=\textrm{diag}\left(1,0,1,0,0,1,0,1\right),\\
&&O_3={1\over 2}\left(\begin{array}{rrrrrrrr}
1 & 0 & 0 & 0 & 0 & 0 & 0 & 1 \\
0 & 1 & 0 & 0 & 0 & 0 & 1 & 0 \\
0 & 0 & 1 & 0 & 0 & 1 & 0 & 0 \\
0 & 0 & 0 & 1 & 1 & 0 & 0 & 0 \\
0 & 0 & 0 & 1 & 1 & 0 & 0 & 0 \\
0 & 0 & 1 & 0 & 0 & 1 & 0 & 0 \\
0 & 1 & 0 & 0 & 0 & 0 & 1 & 0 \\
1 & 0 & 0 & 0 & 0 & 0 & 0 & 1
\end{array}\right)   \label{2001-principle-e13}
\end{eqnarray}
$O_1$ distinguishes the first four states from the second four
and thus induces a partition
(we abbreviate $|\psi_j\rangle $ by $j$)
$\{
\{1 ,2 ,3 , 4 \},
\{5 ,6 ,7 , 8 \}
\}$
of the GHZ-basis states.
$O_2$ distinguishes ${1,2,5,6}$ from ${3,4,7,8}$
and thus induces a partition
$\{
\{1 ,2 ,5 , 6 \},
\{3 ,4 ,7 , 8 \}
\}$
of the GHZ-basis states.
$O_3$ identifies the relative phases
and thus induces a partition
$\{
\{1 ,3 ,5 , 7 \},
\{2 ,4 ,6 , 8 \}
\}$
of the GHZ-basis states.
The three matrices are mutually commutative.
Their combined effect is an atomic partition of the set of base states
$\{
\{1\} ,\{2\} ,\{3\} , \{4 \},
\{5\} ,\{6\} ,\{7\} , \{8 \}
\}$
which is the formal analogue of the experimental sieve.
It is obtained by a successive application of experiments,
represented by the intersection
of each partition element of $O_i$ with all the other ones.
Notice also that\footnote{Recall that
$\sigma_{1x}\equiv \sigma_x\otimes 1_4$,
$\sigma_{2x}\equiv 1_2 \otimes \sigma_x\otimes 1_2$,
$\sigma_{3x}\equiv 1_4\otimes \sigma_x$.
The tensor product $\otimes $ of two second degree tensors $a$, $b$ representable by two
$n\times n$ and $m\times m$ matrices
whose components are $a_{ij}$ and $b_{k,l}$ can be represented by an $nm\times nm$  matrix
$$(a\otimes b)_{s,t}=
    a_{\lceil s/m\rceil , \lceil t/m\rceil }
      b_{s - \lfloor (s - 1)/m\rfloor m\; ,\;
          t - \lfloor (t - 1)/m\rfloor m}, \quad {s,t= 1,\ldots, nm},
$$
where $\lceil x\rceil$ stands for the smallest integer greater than or equal to $x$,
and $\lfloor x\rfloor$ stands for the greatest integer less than or equal to $x$, respectively.
}
$
O_3={1\over 2}\left(1+\sigma_{1x}\sigma_{2x}\sigma_{3x} \right)
$.

The propositions associated with these operators are as follows.
$O_1$ corresponds to the statement that
``the  spin of the first and the  spin of the second particle are the same in the
$z$-direction.''
$O_2$ corresponds to the statement that
``the  spin of the first and the  spin of the third particle are the same in the
$z$-direction.''
$O_3$  corresponds to the statement that
``an even number of spins
is pointing down when measured in $x$-direction.''
This result can be generalized to the case of $n$ particles in a straightforward manner.
Thereby, $n-1$ propositions characterize the relative spins of the particles,
and also the $n$'th proposition is the same as above.


Another, equivalent but permutated set of propositions was proposed by
Cereceda \cite{Cereceda,mermin}:
$
T_1={1\over2}\left(1+\sigma_{1x}\sigma_{2y}\sigma_{3y}\right)
$, $
T_2={1\over2}\left(1+\sigma_{1y}\sigma_{2x}\sigma_{3y}\right)
$, $
T_3={1\over2}\left(1+\sigma_{1y}\sigma_{2y}\sigma_{3x}\right)
$.
Since the transformation matrix $U^{\textrm{GHZ}}$ remains the same,
the projection operators differ from the previous ones
(\ref{2001-principle-e11})--(\ref{2001-principle-e13})
by a permutation of the propositional system
(\ref{2001-principle-e1})--(\ref{2001-principle-e3})
yielding equivalent, though not identical propositions.
This can be explicitly seen by taking row permutations of the operators
(\ref{2001-principle-e1})---(\ref{2001-principle-e3})
\begin{eqnarray}
O_1&=&\textrm{diag}\left(0,1,1,0,1,0,0,1\right), \label{2001-principle-e1c}\\
O_2&=&\textrm{diag}\left(0,1,1,0,0,1,1,0\right),  \\
O_3&=&\textrm{diag}\left(0,1,0,1,1,0,1,0\right), \label{2001-principle-e3c}
\end{eqnarray}
such that $T_i=U^{\textrm{GHZ}} O_i {U^{\textrm{GHZ}}}^{-1}$, $i=1,2,3$.
The physical interpretations of $T_i$   are
as follows \cite{Cereceda}.
$T_1$ corresponds to the statement that
``the product of the spin of particles 1, 2, and 3 along the axes $x$, $y$, and $y$, respectively is equal to $1$.''
$T_2$ corresponds to the statement that
``the product of the spin of particles 1, 2, and 3 along the axes $y$, $x$, and $y$, respectively is equal to $1$.''
$T_3$  corresponds to the statement that
``the product of the spin of particles 1, 2, and 3 along the axes $y$, $y$, and $x$, respectively is equal to $1$.''
The associated GHZ-base state partitions are
$\{
\{1 ,4 ,6 , 7 \},
\{2 ,3 ,5 , 8 \}
\}$ for $O_1$,
$\{
\{1 ,4 ,5 , 8 \},
\{2 ,3 ,6 , 7 \}
\}$ for $O_2$, and
$\{
\{1 ,3 ,6 , 8 \},
\{2 ,4 ,5 , 7 \}
\}$ for $O_3$, respectively.

Another set of orthonormal base states of eightdimensional Hilbert space contains the
W-state introduced in \cite{zeil-97} and discussed in \cite{dvc-2000}.

\begin{eqnarray}
&&|\phi_1\rangle=  |+++\rangle \nonumber \\
&&|\phi_2\rangle=\frac{1}{\sqrt{3}} \left( |++-\rangle+|+-+\rangle +|-++\rangle \right) \nonumber \\
&&|\phi_3\rangle=\frac{1}{\sqrt{2}} \left( -|++-\rangle+|-++\rangle \right) \nonumber \\
&&|\phi_4\rangle=\frac{1}{\sqrt{6}} \left( -|++-\rangle+2 |+-+\rangle -|-++\rangle \right) \nonumber \\
&&|\phi_5\rangle=\frac{1}{\sqrt{3}} \left(  |+--\rangle+             |-+-\rangle +          |--+\rangle \right) \nonumber \\
&&|\phi_6\rangle=\frac{1}{\sqrt{2}} \left( -|+--\rangle+|--+\rangle \right) \nonumber \\
&&|\phi_7\rangle=\frac{1}{\sqrt{6}} \left( -|+--\rangle+ 2|-+-\rangle -|--+\rangle \right) \nonumber \\
&&|\phi_8\rangle=  |---\rangle \nonumber \\
\end{eqnarray}

The unitary transformation $U^{\textrm{W}}$ is given by
\begin{equation}
U^{\textrm{W}}= \left(
\begin{array}{rrrrrrrr}
 1 & 0 & 0 & 0 & 0 & 0 & 0 & 0 \\
0 & \frac{1}{{\sqrt{3}}} & -\frac{1}{{\sqrt{2}}} & - \frac{1}{{\sqrt{6}}} & 0 & 0 & 0 & 0\\
0 & \frac{1}{{\sqrt{3}}} & 0 & \frac{2}{{\sqrt{6}}}                       & 0 & 0 & 0 & 0 \\
0 & \frac{1}{{\sqrt{3}}} &  \frac{1}{{\sqrt{2}}} & -\frac{1}{{\sqrt{6}}} & 0 & 0 & 0 & 0\\
0 & 0 & 0 & 0 &\frac{1}{{\sqrt{3}}} & -\frac{1}{{\sqrt{2}}} & - \frac{1}{{\sqrt{6}}}                                                                        & 0\\
0 & 0 & 0 & 0 &\frac{1}{{\sqrt{3}}} & 0 & \frac{2}{{\sqrt{6}}}                                                                                             & 0\\
0 & 0 & 0 & 0 &\frac{1}{{\sqrt{3}}} &  \frac{1}{{\sqrt{2}}} & -\frac{1}{{\sqrt{6}}}                                                                        & 0\\
0 & 0 & 0 & 0 & 0 & 0 & 0 & 1 \\
\end{array}  \right).
\end{equation}
The corresponding projection operators are
$
O_1=U^{\textrm{W}}O_1^e{U^{\textrm{W}}}^\dagger =\textrm{diag}\left(1,1,1,1,0,0,0,0\right)
$,
$
O_2=U^{\textrm{W}}O_2^e{U^{\textrm{W}}}^\dagger
$,
$
O_3=U^{\textrm{W}} O_3^e{U^{\textrm{W}}}^\dagger
$.


Finally we would like to mention the orthonormal basis resulting from
the orthogonal (and thus unitary) transformation $U^{\textrm{W}}$
\begin{equation}
U^{\textrm{W}}= {1\over 2\sqrt{2}}\left(
\begin{array}{rrrrrrrr}
{1}& {1}& {1}& {1}& {2}& 0& 0&  0 \\
{1}& {1}& {1}& -{1}& -{1}& {1}& -{1}& -{1}  \\
{1}& {1}& -{1}& {1}& -{1}& -{1}& {1}& -{1}  \\
{1}& {1}& -{1}& -{1}& 0& 0& 0& {2}  \\
{1}& -{1}& {1}& {1}& -{1}& -{1}& -{1}& {1}  \\
{1}& -{1}& {1}& -{1}& 0& 0& {2}& 0  \\
{1}& -{1}& -{1}& {1}&  0& {2}&  0&  0 \\
{1}& -{1}& -{1}& -{1}& {1}& -{1}& -{1}& -{1}  \end{array}
\right) .
\end{equation}
It contains for basis states in which
all elements of the standard  orthogonal basis occur equally weighted
(the remaining elements can be obtained by a Gram-Schmidt orthogonalization process).
\begin{eqnarray}
&&|\rho_1\rangle=2\sqrt{2}
\left(   |+++\rangle + |++-\rangle + |+-+\rangle + |-++\rangle + |+--\rangle + |-+-\rangle + |--+\rangle + |---\rangle \right)\nonumber\\
&&|\rho_2\rangle=2\sqrt{2}
\left(   |+++\rangle + |++-\rangle + |+-+\rangle + |-++\rangle - |+--\rangle - |-+-\rangle - |--+\rangle - |---\rangle \right)\nonumber\\
&&|\rho_3\rangle= 2\sqrt{2}
\left(  |+++\rangle + |++-\rangle - |+-+\rangle - |-++\rangle + |+--\rangle + |-+-\rangle - |--+\rangle - |---\rangle \right)\nonumber\\
&&|\rho_4\rangle=2\sqrt{2}
\left(   |+++\rangle - |++-\rangle + |+-+\rangle - |-++\rangle + |+--\rangle - |-+-\rangle + |--+\rangle - |---\rangle \right)\nonumber\\
&&|\rho_4\rangle= 2\sqrt{2}
\left(  |+++\rangle - |++-\rangle + |+-+\rangle - |-++\rangle + |+--\rangle - |-+-\rangle + |--+\rangle - |---\rangle \right)\nonumber
\end{eqnarray}

For the above cases, projection operators $Q_j=U^{\textrm{W}}O_i^e{U^{\textrm{W}}}^\dagger$,
$j=1,2,3$, can be defined, whereby
the operators $O_1,O_2,O_3$ produce partitions
$\{
\{1 ,2 ,3 , 4 \},
\{5 ,6 ,7 , 8 \}
\}$
,
$\{
\{1 ,2 ,5 , 6 \},
\{3 ,4 ,7 , 8 \}
\}$
,
$\{
\{1 ,3 ,5 , 7 \},
\{2 ,4 ,6 , 8 \}
\}$ of the original basis, respectively.
(Any vertical permutation thereof would be equally suitable.)
As before, all $Q_i$ can be given a
direct physical interpretation in terms of ``clicks in a counter'' \cite{rzbb},
but their meaning cannot be expressed in elementary statements.

In summary, we have shown that, given $n$ quantized two-state systems,
 $n$ propositions are enough to
find and separate any
individual pure state
from others of an arbitrary orthogonal basis.
There exist $2^n!$ equivalent sets of $n$ propositions achieving this.
They all differ by permutations from one another.
By considering the simplest case of the standard orthogonal (``Cartesian'') basis,
we have been able to explicitly construct these sets of propositions and their
corresponding projection operators.
Any other orthogonal basis system and the corresponding more general projection operators
can be obtained from this standard orthogonal one
by unitary transformations.
We have explicitly discussed two equivalent solutions
of the ``birthday present puzzle'' for the GHZ-base states and mentioned
the W-state and more general cases.
We conclude that the optimal strategy to single out particular states is in general
based on a measurement of the joint properties of the particles rather than
the properties of the individual particles.

\section*{Acknowledgments}
The authors would like to acknowledge stimulating discussions and suggestions
by {\v{C}}aslav Brukner and Anton Zeilinger.
(Almost needless to say, only the authors should be made responsible
for any misinterpretations and/or errors.)



\begin{thebibliography}{14}
\expandafter\ifx\csname natexlab\endcsname\relax\def\natexlab#1{#1}\fi
\expandafter\ifx\csname bibnamefont\endcsname\relax
  \def\bibnamefont#1{#1}\fi
\expandafter\ifx\csname bibfnamefont\endcsname\relax
  \def\bibfnamefont#1{#1}\fi
\expandafter\ifx\csname citenamefont\endcsname\relax
  \def\citenamefont#1{#1}\fi
\expandafter\ifx\csname url\endcsname\relax
  \def\url#1{\texttt{#1}}\fi
\expandafter\ifx\csname urlprefix\endcsname\relax\def\urlprefix{URL }\fi
\providecommand{\bibinfo}[2]{#2}
\providecommand{\eprint}[2][]{\url{#2}}

\bibitem[{\citenamefont{Zeilinger}(1999)}]{zeil-99}
\bibinfo{author}{\bibfnamefont{A.}~\bibnamefont{Zeilinger}},
  \bibinfo{journal}{Foundations of Physics}
  \textbf{\bibinfo{volume}{29}}(\bibinfo{number}{4}), \bibinfo{pages}{631}
  (\bibinfo{year}{1999}).

\bibitem[{\citenamefont{Birkhoff and von Neumann}(1936)}]{birkhoff-36}
\bibinfo{author}{\bibfnamefont{G.}~\bibnamefont{Birkhoff}} \bibnamefont{and}
  \bibinfo{author}{\bibfnamefont{J.}~\bibnamefont{von Neumann}},
  \bibinfo{journal}{Annals of Mathematics}
  \textbf{\bibinfo{volume}{37}}(\bibinfo{number}{4}), \bibinfo{pages}{823}
  (\bibinfo{year}{1936}).

\bibitem[{\citenamefont{Mackey}(1957)}]{ma-57}
\bibinfo{author}{\bibfnamefont{G.~W.} \bibnamefont{Mackey}},
  \bibinfo{journal}{Amer. Math. Monthly, Supplement}
  \textbf{\bibinfo{volume}{64}}, \bibinfo{pages}{45} (\bibinfo{year}{1957}).

\bibitem[{\citenamefont{Kochen and Specker}(1967)}]{kochen1}
\bibinfo{author}{\bibfnamefont{S.}~\bibnamefont{Kochen}} \bibnamefont{and}
  \bibinfo{author}{\bibfnamefont{E.~P.} \bibnamefont{Specker}},
  \bibinfo{journal}{Journal of Mathematics and Mechanics}
  \textbf{\bibinfo{volume}{17}}(\bibinfo{number}{1}), \bibinfo{pages}{59}
  (\bibinfo{year}{1967}), \bibinfo{note}{reprinted in \cite[pp.
  235--263]{specker-ges}}.

\bibitem[{\citenamefont{Svozil}(1998)}]{svozil-ql}
\bibinfo{author}{\bibfnamefont{K.}~\bibnamefont{Svozil}},
  \emph{\bibinfo{title}{Quantum Logic}} (\bibinfo{publisher}{Springer},
  \bibinfo{address}{Singapore}, \bibinfo{year}{1998}).

\bibitem[{\citenamefont{Cereceda}(2000)}]{Cereceda}
\bibinfo{author}{\bibfnamefont{J.~L.} \bibnamefont{Cereceda}},
  \emph{\bibinfo{title}{Solution of the {B}irthday {P}resent {P}uzzle.
  {P}rivate communication}} (\bibinfo{year}{2000}).

\bibitem[{\citenamefont{Mermin}(1990)}]{mermin}
\bibinfo{author}{\bibfnamefont{N.~D.} \bibnamefont{Mermin}},
  \bibinfo{journal}{Physics Today}
  \textbf{\bibinfo{volume}{43}}(\bibinfo{number}{6}), \bibinfo{pages}{9}
  (\bibinfo{year}{1990}).

\bibitem[{\citenamefont{Zeilinger et~al.}(1997)\citenamefont{Zeilinger, Horne,
  and Greenberger}}]{zeil-97}
\bibinfo{author}{\bibfnamefont{A.}~\bibnamefont{Zeilinger}},
  \bibinfo{author}{\bibfnamefont{M.~A.} \bibnamefont{Horne}}, \bibnamefont{and}
  \bibinfo{author}{\bibfnamefont{D.~M.} \bibnamefont{Greenberger}}, in
  \emph{\bibinfo{booktitle}{NASA Conf. Publ. No. 3135}}
  (\bibinfo{publisher}{National Aeronautics and Space Administration, Code
  NTT}, \bibinfo{address}{Washington, DC}, \bibinfo{year}{1997}).

\bibitem[{\citenamefont{D{\"{u}}r et~al.}(2000)\citenamefont{D{\"{u}}r, Vidal,
  and Cirac}}]{dvc-2000}
\bibinfo{author}{\bibfnamefont{W.}~\bibnamefont{D{\"{u}}r}},
  \bibinfo{author}{\bibfnamefont{G.}~\bibnamefont{Vidal}}, \bibnamefont{and}
  \bibinfo{author}{\bibfnamefont{J.~I.} \bibnamefont{Cirac}},
  \bibinfo{journal}{Physical Review} \textbf{\bibinfo{volume}{A62}},
  \bibinfo{pages}{062314} (\bibinfo{year}{2000}).

\bibitem[{\citenamefont{Reck et~al.}(1994)\citenamefont{Reck, Zeilinger,
  Bernstein, and Bertani}}]{rzbb}
\bibinfo{author}{\bibfnamefont{M.}~\bibnamefont{Reck}},
  \bibinfo{author}{\bibfnamefont{A.}~\bibnamefont{Zeilinger}},
  \bibinfo{author}{\bibfnamefont{H.~J.} \bibnamefont{Bernstein}},
  \bibnamefont{and} \bibinfo{author}{\bibfnamefont{P.}~\bibnamefont{Bertani}},
  \bibinfo{journal}{Physical Review Letters} \textbf{\bibinfo{volume}{73}},
  \bibinfo{pages}{58} (\bibinfo{year}{1994}), \bibinfo{note}{see also
  \cite{murnaghan}}.

\bibitem[{\citenamefont{Specker}(1990)}]{specker-ges}
\bibinfo{author}{\bibfnamefont{E.}~\bibnamefont{Specker}},
  \emph{\bibinfo{title}{Selecta}} (\bibinfo{publisher}{Birkh{\"{a}}user
  Verlag}, \bibinfo{address}{Basel}, \bibinfo{year}{1990}).

\bibitem[{\citenamefont{Murnaghan}(1962)}]{murnaghan}
\bibinfo{author}{\bibfnamefont{F.~D.} \bibnamefont{Murnaghan}},
  \emph{\bibinfo{title}{The Unitary and Rotation Groups}}
  (\bibinfo{publisher}{Spartan Books}, \bibinfo{address}{Washington},
  \bibinfo{year}{1962}).

\bibitem[{\citenamefont{Smith}(1947)}]{smith47}
\bibinfo{author}{\bibfnamefont{C.~A.~B.} \bibnamefont{Smith}},
  \bibinfo{journal}{Mathematical Gazette} \textbf{\bibinfo{volume}{31}},
  \bibinfo{pages}{31} (\bibinfo{year}{1947}).

\bibitem[{\citenamefont{Cairns}(1963)}]{cairns63}
\bibinfo{author}{\bibfnamefont{S.~S.} \bibnamefont{Cairns}},
  \bibinfo{journal}{Amer. Math. Monthly}
  \textbf{\bibinfo{volume}{70}}(\bibinfo{number}{5}), \bibinfo{pages}{136}
  (\bibinfo{year}{1963}).

\end{thebibliography}

\end{document}